# The Metaverse Is Not a Place Apart: Law, Code, and the Recursive Governance of Digital Space

## A Review Essay on Mark Findlay, *Governing the Metaverse: Law, Order and Freedom in Digital Space (Edward Elgar, 2025)*

**Oren Perez**

**Director, Center for Environmental and Climate Law**

**BIU Faculty of Law, *Oren.perez@biu.ac.il***

### Abstract

This review essay examines Mark Findlay's Governing the Metaverse: Law, Order and Freedom in Digital Space. Findlay offers an ambitious and timely account of the metaverse as a social and imaginative space that should be governed for freedom, personhood, community, and resistance to enclosure. The essay argues, however, that the book's two central categories — "the metaverse" and "new law" — remain insufficiently theorised. The book relies on a realspace/virtual distinction that its own analysis repeatedly destabilises. Once digital environments are understood as dependent on physical infrastructures, platform architectures, AI systems, data pipelines, and external legal institutions — and as capable of generating real-world harms for individuals and society — the governance problem is no longer how to devise a separate law for a separate virtual realm. It is how to govern a hybrid socio-technical order in which law, code, platforms, and public oversight recursively interact. The essay further argues that Findlay's account of "new law" does not adequately theorise how normative authority operates across a recursively layered governance architecture in which code, platform rules, and legal oversight interact without any single level exercising decisive control. Drawing on algorithmic constitutionalism, speech-act pluralism, and fuzzy legality, the essay suggests that addressing this architecture requires a jurisprudence capable of reasoning about normative force that is layered, defeasible, and recursively unstable.

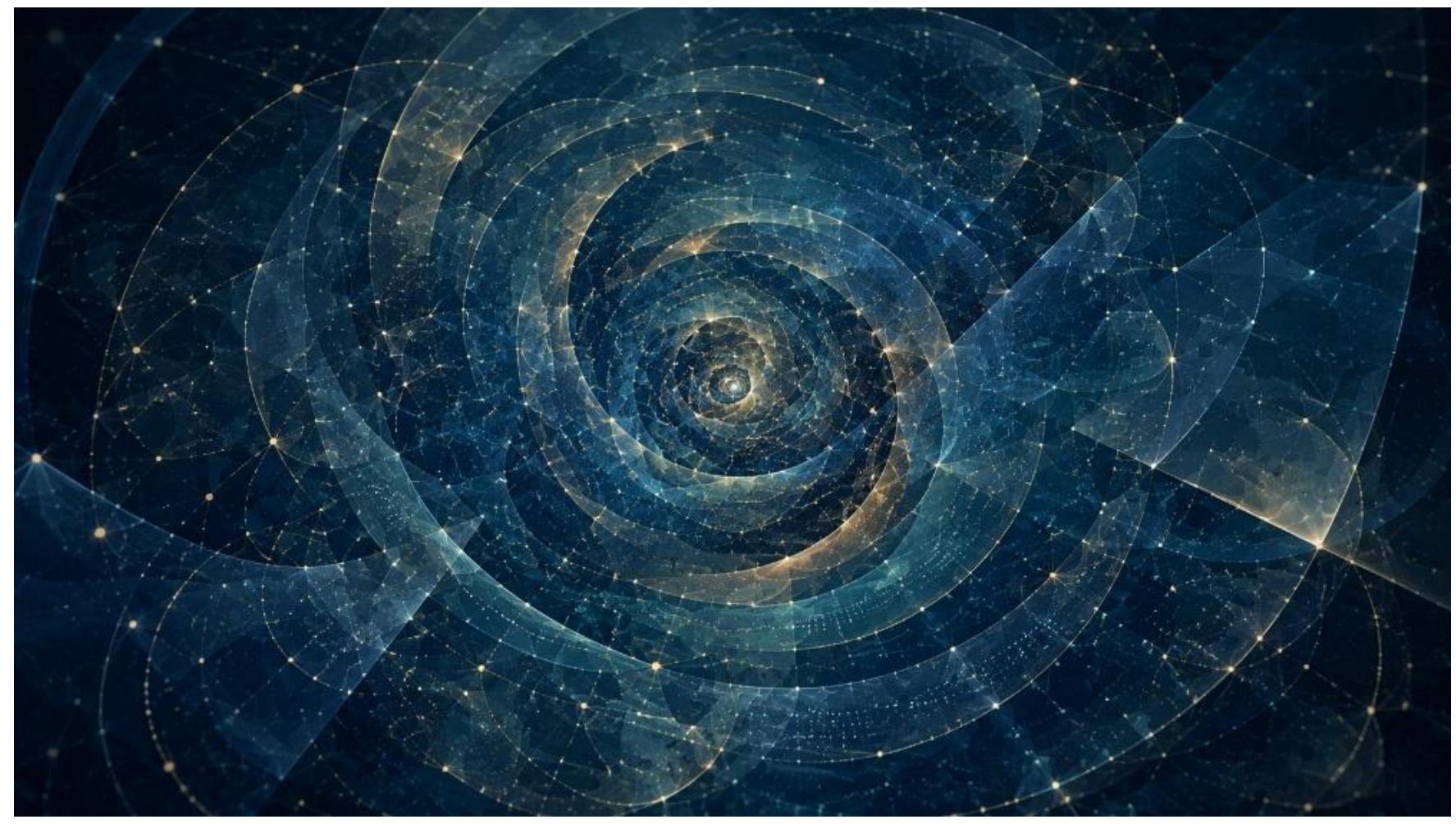

Mark Findlay's *Governing the Metaverse* is part of a large and growing literature that examines various aspects of the metaverse, interrogating how code, norms, markets, and law together determine whether it becomes a realm of freedom or enclosure.[1] Findlay ambitiously and consistently sides with freedom. He seeks to rethink governance for digital space by imagining the metaverse as a social space oriented toward liberty, personhood, and social ordering, one shaped by internal communal needs rather than by external corporate domination. A key theme of his argument is that governing the digital world requires more than transplanting realspace law into the virtual. What is needed, he suggests, is "new law" or "transformed law": a legal form capable of supporting freedom, integrity, community, and trust, rather than the markets and power hierarchies that comprise realspace. Chapter 1 provides the conceptual backbone for the book, which is then elaborated across the later chapters; Chapter 2 examines migration into digital space, using the experience of vulnerable migrant labour to show how digital transition can reproduce exclusion while also enabling new forms of belonging; Chapter 3 develops this claim through the opposition between freedom and enclosure, exploring law and code as competing but potentially complementary grammars of social ordering; Chapter 4 recasts the metaverse as a space of stories and storytelling; Chapter 5 diagnoses the pervasive dependencies generated by digital infrastructures; Chapter 6 develops a more concrete institutional proposal through the idea of digital self-determination; and Chapter 7 attempts to crystallise the book's normative endpoint in the idea of "new law."

The book's normative agenda is highly timely. At a time when AI and platform technologies are concentrated in the hands of a small number of firms with unprecedented global power, Findlay's claim that we need a new law for virtual space that will not simply replicate the hierarchies of realspace is compelling. He is also right to insist that the metaverse is not merely a technical environment but a social and imaginative one. Findlay consistently opposes a vision of digital space as a colonial extension of real-space markets, treating its governance instead as a project of social ordering for liberty and equality, one that will avoid "perpetuating in the virtual the discrimination prevailing from the North World towards the South World in realspace" (p. 22).

One difficulty, however, arises at the level of the book's most basic concept. Although the book is explicitly about governing the metaverse, it does not attempt to develop a clear definition of that concept. Chapter 1, titled "Conceptual locations," would have been the obvious place to provide one. While Findlay insists on demarcating the virtual from realspace, it is not quite clear how this boundary is set. In one place (fn. 2), he notes that because of reservations about the term "metaverse" and its association with big tech, he prefers to use the term "digital

[1] A. Borroni et al., *Metaverses: Reshaping Law, Economy, and Society* (2026); F. Di Porto and O. Pollicino (eds.), *Law and Politics of Virtual Worlds: How NFTs and the Metaverse Are Reshaping Legal Systems* (2026); A. Zhuk, 'Ethical Implications of AI in the Metaverse' (2025) 5 *AI Ethics* 5643; L.B. Rosenberg, 'Regulating the Metaverse: A Blueprint for the Future' in *Extended Reality*, eds. L.T. De Paolis, P. Arpaia, and M. Sacco (2022).

.

universe," adding that "metaverse," "digital universe," and "the virtual" will be used interchangeably against "realspace." Elsewhere, he writes:

> "For the purposes of the analysis to follow, the interpretation of the virtual is both agnostic and pluralist. It is a social place in need of governance for social ordering and where law in a new iteration can stake a claim to this governance outcome. It is a realm into which humans previously reverted in dreaming and is now available through switching on frontier technology and utilising digital communication to create new identities and new communities. In this analysis, understanding a digital universe will not be troubled by technological superstructures or data highways" (p. 15).

The lack of a clearer account of virtual space is not merely a terminological problem; it reflects a conceptual difficulty at the very core of the project. A book about governing the metaverse should be clear about what exactly is being governed. A tautological formulation such as "a social place in need of governance" cannot do that work. This absence matters all the more because, as recent literature shows, the definition of the metaverse is itself highly contested. Ritterbusch and Teichmann's systematic review proposes a relatively narrow definition, describing the metaverse as "a three-dimensional online environment in which users represented by avatars interact with each other in virtual spaces decoupled from the real physical world."[2] Boztas et al. broaden this picture by emphasizing the socio-economic, experiential, and technological dimensions of the metaverse,[3] while Ben Chester Cheong approaches the metaverse through the concept of the avatar and the legal problems that follow from it.[4] The point is not that Findlay had to resolve this debate definitively. It is that he could not simply bypass it.

A further difficulty, and in some respects a more fundamental one, concerns the book's tendency to treat "realspace" and "the virtual" as more sharply distinct than its own analysis actually permits. In Chapter 1 he acknowledges that the divide is not as sharp as it might seem: "isn't it the daily experience that we flow between both seamlessly, and it is the ambivalent merging of *space* that is attractive?" (p. 12). Chapter 2 makes the point even more directly in the context of digital migration: "The borders between physical and virtual spaces are much more permeable than the jurisdictional barriers to freedom of movement in realspace migration" (p. 43). In Chapter 3, Findlay similarly presents the metaverse as a collective sensemaking process and treats law and code as overlapping languages of social ordering rather than wholly separate domains.

Yet these acknowledgments are not fully carried through. The book continues to rely on the real/virtual distinction as the organising axis of its governance argument, without seriously engaging an increasingly important alternative: that the social transformation we are witnessing is not dualistic, with the metaverse and realspace existing as two self-standing spheres, but hybrid, involving an emerging order in which physical and virtual realities are deeply entangled through AI, the Internet of Things, digital twins, and platform infrastructures. In that setting, the virtual does not merely mirror the real, nor does it stand apart from it; rather, it is constitutively intertwined with it. Virtual environments can generate risks in the physical world, both for individuals and for society, while freedom within virtual environments depends on real-world legal protections, infrastructures, and institutional frameworks.[5]

[2] G.D. Ritterbusch and M.R. Teichmann, 'Defining the metaverse: A systematic literature review' (2023) 11 *IEEE Access* 12368, 12372.

[3] C. Boztas et al., 'Metaverse architectures: A comprehensive systematic review of definitions and frameworks' (2025) 17(7) *Future Internet* 283: "The metaverse is a three-dimensional virtual environment that encompasses a vast universe of interconnected digital realms, enabled by advanced technologies, that exist parallel to the physical world, where users are represented by digital representatives, providing an immersive social, economic and interactive experience that transcends traditional internet applications by blurring the boundaries between real and virtual worlds".

[4] B.C. Cheong, 'Avatars in the metaverse: potential legal issues and remedies' (2022) 3(2) *International Cybersecurity Law Review* 467.

This point matters because it weakens a core organising assumption of the book. Once one takes seriously the extent to which virtual worlds depend on physical infrastructures and can produce physical consequences, the question is no longer whether a distinct law for the metaverse can be devised apart from realspace. Rather, the central problem is how law and governance should address a hybrid socio-technical order in which physical and virtual relations are mutually constitutive. The concept of complementary pairs helps clarify what this means: virtual~real, law~code, freedom~enclosure are not binary oppositions between self-standing domains but relational distinctions in which each term is constitutively shaped by its counterpart.[6] On that view, governance of the metaverse cannot be approached as governance of a separate realm. It necessarily includes the regulation of the real-world infrastructures, firms, data pipelines, sensor systems, and institutional arrangements that make metaverse activity possible, facilitate the freedoms available within it, and govern the risks it creates.

To be sure, Findlay does acknowledge these dependencies, and Chapter 5 is one of the book's strongest chapters. Its contribution lies not simply in observing that digital life depends on AI-assisted technologies, but in showing that such dependencies are structured by cultures of convenience, unequal access to information, and power asymmetries. Findlay moves instructively from relatively mundane forms of overreliance, such as dependence on Google Maps for navigation, to more harmful scenarios, including discriminatory uses of algorithmic systems in employment. He also argues, persuasively, that any meaningful response to AI–human dependency must begin with information-sharing: data subjects need to know the nature and consequences of the dependencies in which they are implicated if they are to exercise any meaningful form of self-determination and, with it, any capacity to contest or regulate those dependencies. The chapter's treatment of trust is equally significant. Because AI–human dependencies expose and often deepen human vulnerability, the question of whether digital technologies can be trusted becomes a key regulatory concern. Yet the book does not fully absorb the implications of this insight. If freedom in virtual space depends on real-world infrastructures, then the governing problem is not one of two separate spaces set against one another, but of a deeply entangled order. Chapter 5 repeatedly moves in this direction, but the book stops short of incorporating that insight into its wider conceptual framework.

[5] X. Wang et al., 'Integration of sensing, communication, and computing for metaverse: A survey' (2024) 56(10) *ACM Computing Surveys* 1; K. Li et al., 'When internet of things meets metaverse: Convergence of physical and cyber worlds' (2023) 10(5) *IEEE Internet of Things Journal* 4148; N. Kshetri, 'Privacy and cybersecurity issues facing the metaverse: An analysis of technological and institutional factors' (2026) 50 *Telecommunications Policy* 103140.

[6] For the concept of complementary pairs and its application to hybrid institutional orders, see O. Perez, 'The hybrid legal-scientific dynamic of transnational scientific institutions' (2015) 26(2) *European Journal of International Law* 391.

Chapter 4 can be read as Findlay's effort to ease the tension generated by the duality thesis. He notes that storytelling "is presented to assist a more fluid appreciation of virtual life experiences and breaking down unhelpful dualities of actual and imagined as they shape digital personhood and thereby require social ordering (governance)" (p. 105). The chapter accordingly proposes storytelling as a way of understanding the metaverse as a realm of imaginative possibility. Findlay emphasizes the capacity of storytelling to explore existential and unconscious life experiences that existed within and before the digital, and thereby to illuminate the linkages between the digital and real universes. He connects storytelling to the idea of semiotic democracy, understood as a participatory process in which both the storyteller and her audience combine in an experience that empowers all who participate through active engagement in the production of meanings. This is a genuinely imaginative effort to reconnect legal analysis to questions of narrative, identity, memory, and participation.

However, the book's reliance on storytelling as a method for analysing the governance dilemmas of the metaverse also reveals a methodological vulnerability. Findlay does not fully confront the limits of storytelling as a method of legal analysis. Law is not merely a narrative practice or an "artefact of stories and storytelling" (p. 107). It is an institutionalised order oriented towards making binding normative judgments, underpinned by the fundamental principles of legality and validity. There is, accordingly, only so much that storytelling, however suggestive, can contribute to the analysis of a phenomenon of this kind.

This concern is deepened by the most distinctive methodological feature of the book: its sustained use of Lewis Carroll's *Alice's Adventures in Wonderland* as a governing metaphor. References to Alice, the Cheshire Cat, the Queen of Hearts, and the logic of Wonderland pervade the text, and Chapter 1 explicitly develops this as a form of "metaphor methodology." The device has undeniable imaginative appeal. Wonderland's inverted logic, where sentences precede verdicts, roads lead nowhere, and identity is perpetually unstable, does capture something real about the disorientation that legal and regulatory thinking experiences when confronted with virtual space. Yet, applied so pervasively, the metaphor eventually overreaches. The difficulties of governing the metaverse, including platform power, data exploitation, algorithmic discrimination, and jurisdictional fragmentation, exceed what the Wonderland metaphor can illuminate. They require a techno-institutional account, not only narrative imagination.

These theoretical difficulties are carried forward into Chapter 7, where they emerge most fully in Findlay's explicit account of "new law." New law, he argues, does not derive its authority from compulsion, does not operate primarily through command-and-control rules, and is not trapped in the dysfunctional paradox of realspace law, "where justice is declared as for all and enclosed for the few via the commodification of legal services to benefit those who pay its rent" (p. 200). On this account, new law is a communal resource, driven by a reflexive rather than a coercive logic, and directed toward resisting enclosure and enabling freedom. These are powerful normative aspirations, and Chapter 7 situates them within a demanding theoretical framework, drawing on Teubner's reflexive law, Benjamin's critique of law's violence, and self-subversive justice to imagine a legal order beyond coercion and market enclosure.

Yet the chapter also exposes the depth of the book's jurisprudential difficulty. It does not explain in what sense the normative ordering it describes remains law. In particular, it overlooks a fundamental feature of law: law's constitutive claim to authority, reflected in its capacity to make binding determinations about rights and duties, normative validity, and juridical competence. Once that claim to authority is substantially loosened, it becomes questionable whether what remains can still be described as "new law" in any meaningful jurisprudential sense. Societies may, of course, generate order without law, but the scale and structure of the problems raised by digital freedom, platform power, and infrastructural dependency make it difficult to see how communal bonding and mutualised interests alone could supply an adequate governance response. The chapter therefore offers a powerful critique of realspace law, but it does not provide a convincing account of how "new law" would be structured, how it would claim authority, or how it would address the governance challenges the book so effectively diagnoses.

A further dimension of this jurisprudential difficulty appears once digital governance is approached through the lens of algorithmic constitutionalism. From that perspective, the book's account of "new law" leaves underexplored the interaction between external legal authority and the internal rule structures of digital systems. Digital spaces are not governed only by legal rules imposed from outside; they are also governed from within by code, platform architectures, automated enforcement systems, and embedded design choices that perform constitution-like functions.[7] One of the key questions in contemporary digital governance is therefore not only whether external law can regulate platforms and AI systems, but also how these internal rule structures can be constrained, justified, and made accountable. The idea of algorithmic constitutionalism is useful here because it treats digital governance as a problem of constitutional design within algorithmic systems themselves: a layered architecture of object-level rules and meta-level constraints, combined with mechanisms of algorithmic meta-reasoning and corrective deliberation through which external actors can contest and seek the revision of algorithmic decisions.[8] This framework highlights a point that Findlay's account of "new law" leaves underdeveloped: if digital spaces are governed by code, platform rules, automated enforcement, and embedded decision procedures, then the question is not only what law should do to the metaverse from outside, but how constitutional constraints can be built into the architecture of digital governance from within. Anthropic's constitutional AI approach to Claude provides a contemporary illustration of this challenge: it seeks to guide model behaviour through higher-order principles functioning as internal normative constraints. But the deeper issue is how such internal constitutional structures relate to external legal oversight and democratic contestation.

Algorithmic constitutionalism also clarifies why Findlay's contrast between realspace law and emancipatory virtual ordering is too stark. Internal constitutional design cannot replace external law, but neither can it be reduced to ordinary private ordering. It occupies an intermediate position between code, platform governance, public regulation, and constitutional theory. This is precisely the space in which many contemporary digital governance problems now arise. The challenge is therefore not to choose between state law and communal virtual self-ordering, but to understand how algorithmic systems, platform institutions, public law, and deliberative oversight interact in hybrid constitutional arrangements. Seen in this light, the book's account of "new law" is suggestive but incomplete: it gestures toward reflexivity, participation, and freedom, but does not specify the institutional or technical mechanisms through which these values could be embedded and made durable inside digital systems that are already structured by automated decision-making and platform power.

[7] D. Deepak and B. Manski, 'Coding the future: Digital technologists and the constitution of the next system' (2025) 10 *Frontiers in Sociology* 1362848.
[8] O. Perez and N. Wimer, 'Algorithmic constitutionalism' (2023) 30(2) *Indiana Journal of Global Legal Studies* 81.

The structural complexity of this hybrid constitutional order calls for analysis that goes beyond both institutional description and conventional jurisprudential categories. Its governance architecture is not simply hierarchical but recursive: each level can revise, contest, or reinterpret the normative outputs of the others, without any single level exercising decisive control over the structure as a whole. Ground-level code and platform rules establish and enforce first-order norms; internal meta-level norms oversee and qualify them; and external legal oversight functions as a partial meta-governor, reviewing, revising, or invalidating governance decisions across layers. Yet those interventions translate only imperfectly back into ground-level code, which then generates further normative consequences that may themselves be contested or defeated. The cycle does not produce full closure. Each translation between levels may transform, attenuate, or displace normative content, and no level wholly determines any other. This recursive structure matters jurisprudentially because normative force is not fixed at a single point in the architecture; it is produced, qualified, translated, and contested across layers. Conventional jurisprudential frameworks struggle to capture this configuration: one in which normative conventions are multiple and potentially conflicting across layers, and normative force is graded and defeasible rather than binary — a terrain that speech-act pluralism and fuzzy legality help theorise.[9]

Algorithmic constitutionalism therefore calls for a jurisprudence of defeasible normative force — one that takes seriously the autonomy of each layer and the cyclical nature of their relations, and that supplies analytical tools for reasoning about governance outcomes that are structurally unstable and open to revision elsewhere within the architecture.

In the end, the book's difficulty lies not merely in its occasional abstraction or limited practical elaboration. More fundamentally, it turns on two central categories, "the metaverse" and "new law," that remain insufficiently theorised, and whose conceptual weaknesses interact in ways that compromise the book's overall argument. In the case of the metaverse, the problem is not only definitional indeterminacy, but also the book's uneven engagement with the possibility of a hybrid, rather than dual, world: one in which virtual and realspace relations are deeply entangled rather than organised as separate spheres. Chapters 3 and 6 contain important resources for such an account, especially in their treatment of law and code as overlapping ordering languages, and of digital self-determination as a collaborative governance model, but these insights are not fully integrated into the book's wider framework. The jurisprudential gap in the account of "new law" is equally serious. Once digital governance is understood as a recursively unstable normative order whose instability is structural rather than contingent, reflexivity and communal self-ordering cannot do all the work that Findlay assigns to them. What is required is a jurisprudence capable of addressing the layered, defeasible, and recursively unstable normative configuration of digital governance — a demand that Findlay's account of "new law," for all its theoretical ambition, does not meet. The book's central ambition therefore remains compelling but under-realised in both institutional and jurisprudential terms. The final chapter, despite its theoretical depth, gestures toward reflexivity and participation without addressing the cyclical and distributed features of the emergent architecture of digital governance. Ultimately, the gap between the book's theoretical reflections and the practical dilemmas it identifies is not bridged by the framework it provides. Yet if the book falls short as a theory of digital law, it nevertheless succeeds in making the need for such a theory difficult to ignore. Its contribution lies in bringing freedom, dependency, and enclosure into view as problems of digital legal imagination, and in resisting the assumption that virtual space must simply reproduce the hierarchies of realspace. These are not small achievements. The book is therefore best read not as a fully worked-out theory of digital law, but as a provocative, imaginative, and often illuminating intervention that makes the need for such a theory more visible. On those terms, it is a valuable addition to an emerging field that still has too few works of comparable ambition.

[9] O. Perez, 'Tolerance of Incoherence in Law, Graded Speech Acts and Illocutionary Pluralism' (2020) 26 *Legal Theory* 214, 237; O. Perez, 'Fuzzy Law: A Theory of Quasi-Legal Systems' (2015) 28 *Canadian Journal of Law and Jurisprudence* 343.